\documentclass{aa}

\usepackage{amsmath}
\usepackage{siunitx}
\usepackage{xfrac}
\usepackage{natbib}
\usepackage{xcolor}
\usepackage{graphicx}
\usepackage{booktabs}
\usepackage{csquotes}
\usepackage{subcaption}
\usepackage{multirow,array}
\makeatletter
  \renewcommand*\aa@pageof{, page \thepage{} of \pageref*{LastPage}}
\makeatother
\usepackage{hyperref}            

\hypersetup{pdfpagemode = {UseNone},
            pdftitle = {1to1},
            pdfcreator = {\LaTeX},
            pdfproducer = {pdfeTeX-0.\the\pdftexversion\pdftexrevision},
            pdfauthor = {Wilhelm Kley, Daniel Thun, Anna Penzlin},
            pdfsubject = {},
            pdfview = {FitH},
            pdfstartview = {FitH},
            colorlinks = {true},
            linkcolor = [rgb]{0,0.35,0.7},
            citecolor = [rgb]{0,0.35,0.7},
            filecolor = [rgb]{0.61,0,0},
            urlcolor = [rgb]{0.61,0,0},
           }

\usepackage[normalem]{ulem}
           
\usepackage{bm}

\colorlet{darkgreen}{green!60!black}

\colorlet{darkpink}{pink!50!black}

\newcommand{\beq}{\begin{equation}}
\newcommand{\eeq}{\end{equation}}

\begin{document}

\title{1:1 orbital resonance of circumbinary planets}

\author{Anna~B.~T. Penzlin,
	Sareh Ataiee \and
        Wilhelm Kley }

\institute{
Institut f\"ur Astronomie und Astrophysik, Universität T\"ubingen,
Auf der Morgenstelle 10, D-72076 T\"ubingen, Germany\\
\email{\{anna.penzlin, sareh.ataiee, wilhelm.kley\}@uni-tuebingen.de}\\
}

\date{}

\abstract
{The recent detection of the third planet in Kepler-47 has shown that binary stars can host several planets in circumbinary orbits.
To understand the evolution of such systems we have performed two-dimensional hydrodynamic simulations of the circumbinary disc with
two embedded planets for several Kepler systems. In two cases, Kepler-47 and -413, the planets are captured in a 1:1 mean-motion resonance
at the planet parking position near the inner edge of the disc. 
The orbits are fully aligned, have mean eccentricities of about 0.25 to 0.30,
and the planets are entangled in a horseshoe type of motion. 
Subsequent n-body simulations without the disc show that the configurations are stable.
Our results point to the existence of a new class of stable resonant orbits around binary stars.
It remains to be seen if such orbits exist in reality.
}

\keywords{
          Hydrodynamics --
          Binaries: general --
          Planets and satellites: resonance --
         }

\maketitle

\section{Introduction}\label{sec:intro}

By now about a dozen planets orbiting around binary star systems have been detected by the Kepler space mission.
All of these are single-planet systems, except Kepler-47 whose third planet was recently detected \citep{2019Kepler47}, 
demonstrating that circumbinary systems can host multiple planets.
This raises the question about their formation, evolution and stability. 

Following previous approaches we assume that the planets formed
further away from the central binary in more quiet disc regions and then migrated inwards where they
are parked at the cavity edge. Using hydrodynamical disc simulations, we modelled the migration of multiple
planets in 6 of the known Kepler circumbinary systems and discovered stable coorbital resonant configurations in Kepler-47  and -413.

Thus far, 1:1 resonances have not yet been observed in exoplanet systems but only for minor Solar System objects.
Well-knowns are the population of Trojans of Jupiter and Saturn, and the moons Janus and Epimetheus that orbit Saturn,
which have similar mass and are on mutual horseshoe orbits \citep{1983Icar...53..431Y}.
For the latter, \citet{2019Janus} have recently presented a formation mechanism involving migration in a circumplanetary disc.
Relating to exoplanets, formation of 1:1 resonance objects can proceed for example in-situ \citep{2007A&A...463..359B}, or through a
convergent migration process \citep{2006A&A...450..833C}. In the first scenario the planets begin on tadpole orbits,
in the latter case they start out with horseshoe motion that later turn into tadpole as well \citep{2006A&A...450..833C}.

However, continued migration of the resonant pair in a gas disc will lead to an increase in the libration amplitude which can destabilise the system
\citep{2009CresswellNelson}, in particular for more massive planets \citep{2014Pierens}.
Stabilisation of a system can be achieved if the pair is in resonance with a third planet \citep{2009CresswellNelson,2019arXiv190107640L}, whose
presence may also assist in forming the 1:1 resonance in the first place, by providing additional perturbations.
Around a binary star, planets reach a stable parking position near the disc's inner cavity as determined by the binary and the disc parameters.
In case of a successful capture near this position one might 
expect long-term stability for such coorbital configurations as they do not migrate any further.

For the pure n-body case, the dynamics and stability of eccentric coplanar coorbitals around single stars has been studied for the equal planet
mass case in \citet{2018CeMDA.130...24L}. They show that for low mass planets around $10^{-5}$ of the stellar mass, stable horseshoe and
tadpole orbits exist up to an orbital eccentricity of about 0.5.
The stability of 1:1 resonances between low mass planets has been studied in the special case of an equal mass binary
star on a circular orbit by \citet{1989JApA...10..347M}, who found stability only in the zero mass limit of the planets.

Here, we present the disc driven migration of a system of small and equal mass planets in the Kepler-47 and -413 systems.
We show that in both cases stable 1:1 resonant configurations are possible. 
In sect.~\ref{sec:model} we describe our model setup, in sect.~\ref{sec:resonance} we focus on the resulting 1:1 resonances, and in
sect.~\ref{sec:discuss} we discuss out results.
 
\begin{table}
    \centering
    \begin{tabular}{clcc}
        &Kepler & 413 & 47 \\
        \midrule\midrule
        \multirow{4}{*}{\rotatebox[origin=c]{90} {Binary}}
        &$M_\mathrm{A}$ $[M_\sun]$ & 0.82 & 1.04\\
        &$M_\mathrm{B}$ $[M_\sun]$ & 0.54 & 0.36\\
        &$a_\mathrm{bin}$ $[\mathrm{au}]$& 0.10 & 0.08\\
        &$e_\mathrm{bin}$ & 0.04 & 0.02\\
        \midrule
        \midrule
        \multirow{3}{*}{\rotatebox[origin=c]{90} {Cavity}}
        &$a_\mathrm{cav}$ $[a_\mathrm{bin}]$& 5.18 & 5.24\\
        &$e_\mathrm{cav}$ & 0.45 & 0.43\\
        &$T_\mathrm{prec}$ $[T_\mathrm{bin}]$ & 2271 & 2573\\
        \bottomrule
    \end{tabular}
    \caption{Parameters for Kepler-413 and -47, as used in the simulations.
    {\bf Top}: 
    $M_\mathrm{A}$ and $M_\mathrm{B}$ denote the stellar masses of the binary, with
    semi-major axis, $a_\mathrm{bin}$, and eccentricity, $e_\mathrm{bin}$. 
    {\bf Bottom}: Mean equilibrium orbit parameters of the inner cavity of the disc without planets, as obtained by the hydrodynamical evolution.
    $T_\mathrm{prec}$ is the precession period, $a_\mathrm{cav}$ the cavity semi-major axis, and $e_\mathrm{cav}$ its eccentricity.
    \label{tab:disc}
   }
\end{table}
\begin{figure}
    \centering
    \resizebox{\hsize}{!}{\includegraphics{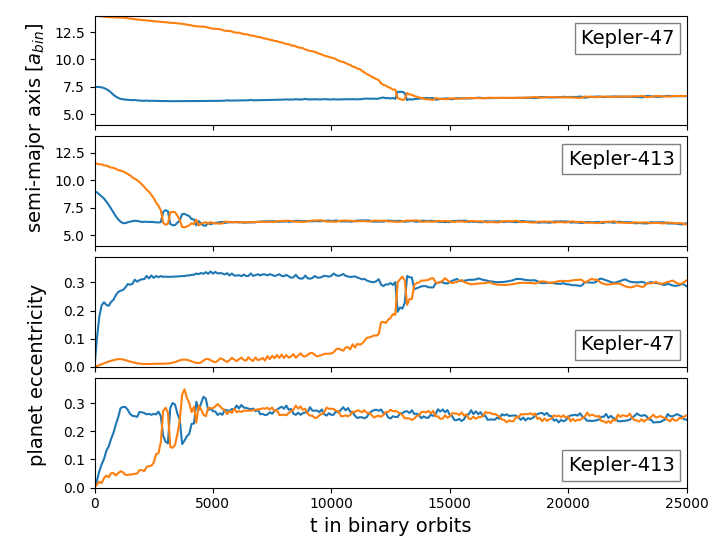}}
    \caption{Time evolution of the orbital elements of the inserted planets in the Kepler-413 and -47 system while embedded in a disc,
     that was simulated for \num{20000}\,$T_\mathrm{bin}$ prior to insertion.
    }
    \label{img:orb}
\end{figure}
\section{Models}\label{sec:model}
To model discs around binary stars, we perform viscous hydrodynamic simulations of thin discs in two dimensions (2D) using
a modified version of the PLUTO-code \citep{2007Pluto} that allows for computation on graphics processing units \citep{2017Thun}.
We use a locally isothermal disc because the final equilibrium structure of circumbinary discs with full radiative cooling and viscous heating is similar to a locally isothermal model using the equilibrium temperature, see \citet{2019Kley}.
They found the disc aspect ratios, $h$, between 0.04 and 0.05, depending on the binary system and disc parameters.
Here, we choose $h = 0.04$, for the viscosity parameter $\alpha=0.001$, and a disc mass of 1\% of the binary mass.
The orbital parameter of the two modelled systems, Kepler-413 and -47, are listed in the upper block of Table~\ref{tab:disc}.

The simulations presented here are performed using the same approach and method as described in \citet{2018Thun} and \citet{2019Kley}. We use cylindrical coordinates centred on the barycentre of the binary and a radially logarithmic grid with $684 \times 584$ cells, 
that stretches from 1 to 40 binary separations, $a_\mathrm{bin}$.
This domain size and resolution has shown to be sufficient for such type of studies \citep{2018Thun}.
For the n-body integrator we use a 4th order Runge-Kutta scheme, which is of sufficient accuracy,
given an average time-step size of $2 \times 10^{-3}\,T_\mathrm{bin}$.

First, we simulate the discs without planets to their convergent states which are reached after about $20\,000$ binary orbits, $T_\mathrm{bin}$,
and then embed the planets.
Simulations of circumbinary discs converge to a disc with a large, eccentric, and precessing inner cavity. 
The mean orbital parameters of the cavities, calculated by fitting the inner disc edge (see \citet{2017Thun} for details),
are given in the lower block of Table~\ref{tab:disc}.
The cavity size and eccentricity is determined by the binary eccentricity and mass ratio, the disc  height and viscosity.
In agreement with \citet{2018Thun}, systems with small $e_\mathrm{bin}$ show very large cavity eccentricities.
In our case we obtain $e_\mathrm{cav}\sim 0.45$, see Table~\ref{tab:disc}. 

Then we add the planets into the disc at different initial locations on circular orbits, and let them migrate. The planets do not accrete mass. 
In case of Kepler-47, we add three planets, initially at 7.5, 14 and 20\,$a_\mathrm{bin}$.
The two inner planets each have a mass of $25\,M_\oplus$ while the outer planet has $5\,M_\oplus$. 
The masses of the planets lie within the error bars of the newly discovered 3 planet system in Kepler-47 \citep{2019Kepler47}.
As the additional outer planet has much smaller mass, and remains further out in the disc, it has little effect on 
dynamical evolution of the two inner planets, and is not discussed here any further.
Even though in Kepler-413 only one planet is observed, we embed two planets at 9 and 11.5 $a_\mathrm{bin}$ each with $0.21\,M_\mathrm{Jup}$
which is identical to the observed planet mass in Kepler-413 \citep{2014ApJ...784...14K}.
\begin{figure}
    \centering
    \resizebox{\hsize}{!}{\includegraphics{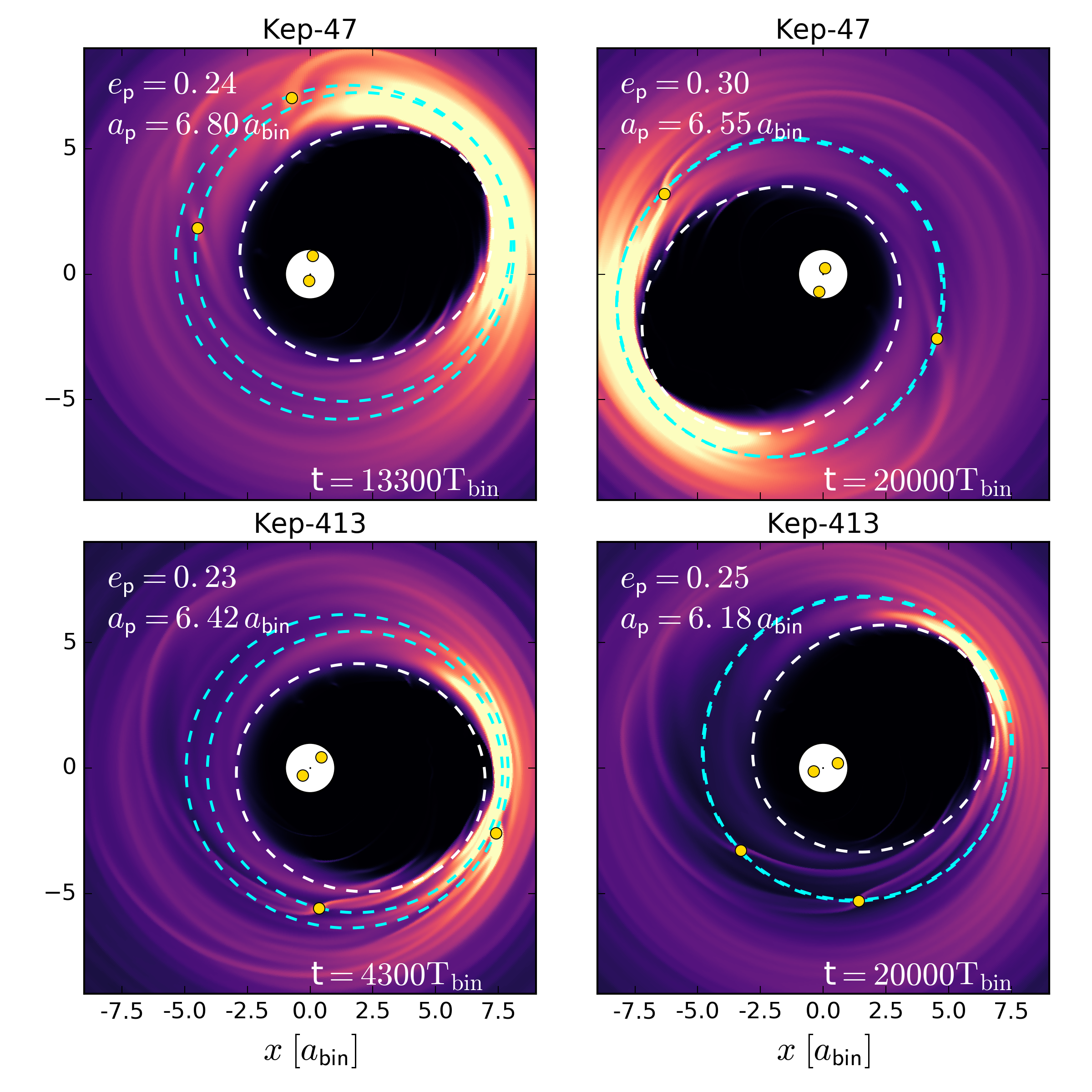}}
    \caption{Density distribution at the time right before the capture (left) and in the coorbital configuration (right) for Kepler-47\,(top) and -413\,(bottom). The white ellipse indicates the location of the cavity in the disc and the light blue ellipses show the orbits of both planets, where the yellow dots mark their positions. 
     The orbital parameters of the outer planet are quoted at the top left corner. 
    }
    \label{img:2d}
\end{figure}
 
\begin{figure}
    \centering
    \resizebox{\hsize}{!}{\includegraphics{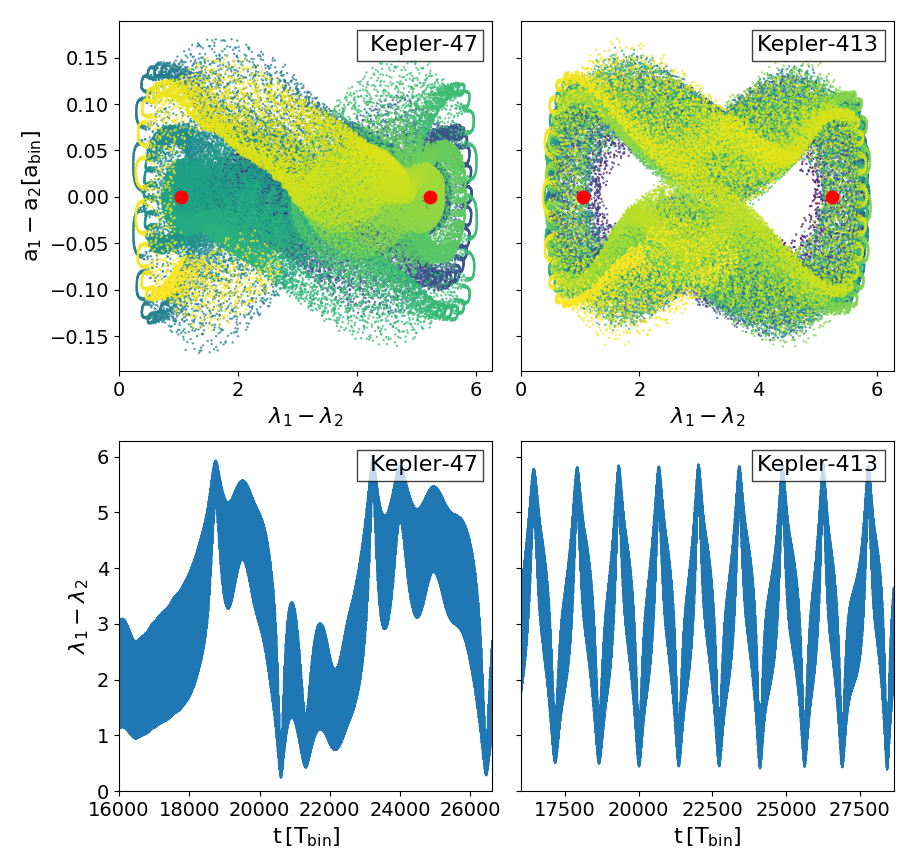}}
\caption{Dynamical structure of the 1:1 resonances while the planets are still embedded in the disc, on the left Kepler-413 and on the right Kepler-47.
{\bf Top}: Semi-major axis difference of the planets against difference in mean longitude. The colour changes linearly from blue to yellow over the time span displayed in the lower panels. The red dots indicate the location of the Lagrange points L4 and L5 in case of circular motion.
{\bf Bottom}: Angular distance between the planets vs. time. 
}
\label{img:dist}
\end{figure}

\section{Forming a stable 1:1 resonance}\label{sec:resonance}
In both cases, the planets migrate toward the stable parking position near the inner cavity of the disc, where is first reached by the inner planet,
blue curves in Fig.\,\ref{img:orb}. Such low mass planets do not open a complete gap, hence their orbits align with the cavity and gain in eccentricity \citep{2019Kley}. 
The second planet, starting further out, takes more time to reach the cavity and acquire eccentricity.
This process takes longer for Kepler-47, as lower mass planets migrate slower, and the outer planet is initialised further away.

As the parking position of planets around binaries is determined by the disc's cavity edge, the incoming second planet reaches the same orbit as the first.
In our cases this leads to capture in 1:1 orbital resonant configurations. As shown in Fig.~\ref{img:orb},
this happens after $\sim 5000\,T_\mathrm{bin}$ for Kepler-413 and $\sim 13\,000\,T_\mathrm{bin}$ for Kepler-47.
During the capture process the two planets show initially larger excursions in semi-major axis and eccentricity.
During this process the planets do not collide with each other, as the minimum distance between them is $\sim 10^{-2}\,a_\mathrm{bin}$.
The excursions are damped by the action of the disc but the orbits remain eccentric.
After the capture, we continue the hydrodynamical simulations for more than $10\,000\,T_\mathrm{bin}$ and the final orbital configuration of the planets remains stable. 
\begin{figure*} 
    \centering
     \resizebox{\hsize}{!}{\includegraphics[width=0.85\textwidth]{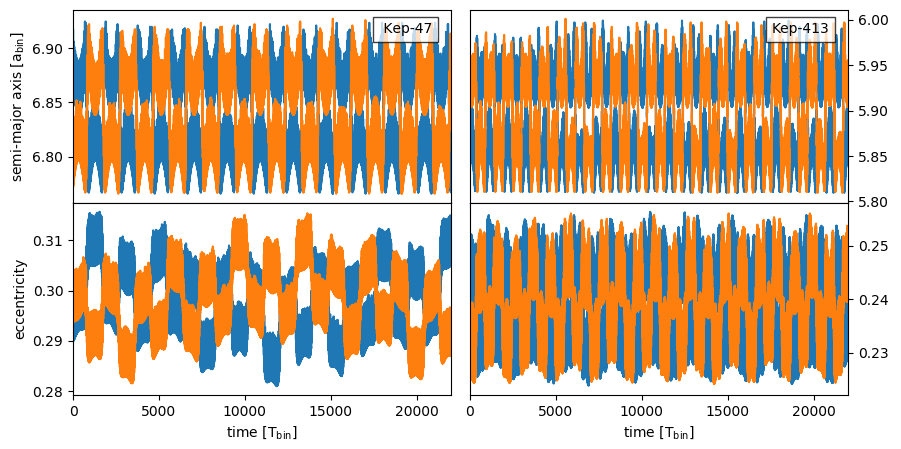}}
    \caption{Evolution of semi-major axes and eccentricities of the two planets orbiting Kepler-413 and -47 for the pure n-body case without disc. 
    }
    \label{img:n-body}
\end{figure*}

The shape of the orbits in relation to the disc's structure in this phase of the simulations is displayed in Fig.~\ref{img:2d},
where snapshots of the configurations are plotted at around the capture time (left panels) and in equilibrium, at $\num{20000}\,T_\mathrm{bin}$ after planet insertion (right panels).
The orbits lie near the disc's inner edge, are more circular than the cavity but still have an eccentricity of $\sim 0.25-0.30$.
They begin to align just before capture, and their precession rate and that of the cavity match exactly as soon as the planets arrive in their stable parking positions.
In the subsequent evolution, the orbits remain fully aligned with a maximum libration angle of $6^\circ$.
For Kepler-413, the precession rate of the cavity is lowered to 1900~$T_\mathrm{bin}$ after introducing the planets.
In Kepler-47, whose planets are less massive than the former, the cavity almost maintains its previous precession rate which is 2600~$T_\mathrm{bin}$. 
The presence of these small planets does not change significantly the cavity size and eccentricity for either system.

The orbital dynamics of the two planets, engaged in the 1:1 resonance while still embedded in the disc, is displayed in Fig.~\ref{img:dist}.
It shows that the difference in mean longitude between the two planets never drops below $20^\circ$. So they never have a close encounter and do not collide, but are rather engulfed in a periodic motion of a 1:1 resonance.
In both cases the planets move on horseshoe orbits around the Lagrange points L4 and L5. In the case of Kepler-413 the motion is quite regular with a libration period of about 1330 $T_\mathrm{bin}$.
In the case of Kepler-47 the motion is more irregular and a planet moves multiple time around one Lagrange point before switching over to the other.
We suspect that the reason for this irregular behaviour is the lower mass of the planets and stronger perturbed disc compared to Kepler-413. In such a condition, the disc perturbation can give larger random acceleration to the planets and send them back and forth to the horseshoe and Lagrange points.

To check the long-term stability of these orbits in the absence of the disc, we extracted the final positions and velocities of the planets and stars,
and used them as the initial condition for a pure n-body simulation containing the two stars with the planets ($n=4$).
In Fig.~\ref{img:n-body}, we show the evolution of semi-major axis and eccentricity over a time span $22\,000\,T_\mathrm{bin}$.
In both systems, the orbits are stable and the planets remain in the 1:1 resonance. 
The orbits retain their previous eccentricity (with disc) and change periodically their positions upon close approach, where
the planet on the inner position switches to the outer and vice versa, as is typical for horseshoe orbits.

\begin{figure}
	\centering
     \resizebox{\hsize}{!}{\includegraphics[width=0.85\textwidth]{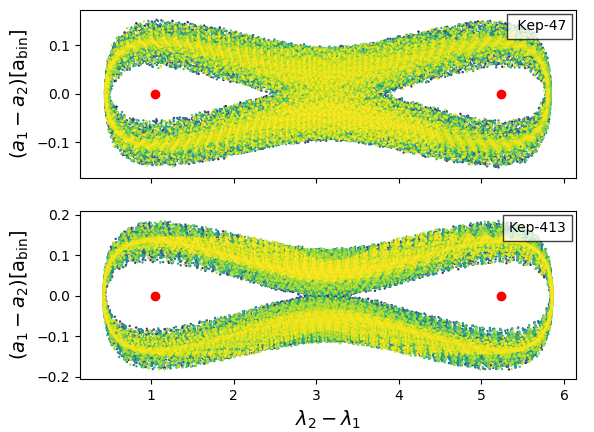}}
\caption{The difference in mean longitude of the two planets for the pure n-body case without a disc. The colour changes linearly with time over 1600 years or $57 \, 900\, T_\mathrm{bin}$ for Kepler-413 and $78\, 500 \, T_\mathrm{bin}$ for Kepler-47.}
    \label{img:angle}    
\end{figure}

In the pure n-body case, the orbits are much more regular than in the case with the disc, and in both systems the planets are on regular, smooth horseshoe orbits, see Fig.~\ref{img:angle}.
Without the disc, the radial excursions of the planets are similar to the case with the disc present.
The libration period is 1920 $T_\mathrm{bin}$ for Kepler-47 and 1055 $T_\mathrm{bin}$ for Kepler-413.
With the disc present, these periods were longer, for example by 300 $T_\mathrm{bin}$ in the case of Kepler-413,
because interaction with the disc influences the orbital properties of the embedded planets.
During one libration period in the n-body simulation, the planets approach each other only up to a mean longitude difference between the planets of about  $25^\circ$, shown in Fig.~\ref{img:angle}, ensuring stability.

Without the disc, the aligned planetary orbits precess around the binary at a slower rate than with a disc present. 
This is expected as the precession of the planets in the disc was aligned with the inner cavity.
In the n-body simulation we find precession periods of $4245 \, T_\mathrm{bin}$ for Kepler-47 and $2275 \, T_\mathrm{bin}$ for Kepler-413.
These periods correspond to that of a single planet with the mean orbital elements of the planet pair.

\section{Discussion \& Conclusion}\label{sec:discuss}

Using a migration scenario for two planets in a circumbinary disc we have shown that low eccentric binary systems like Kepler-47 and -413
are able to host two small, equal-mass planets in a stable coorbital resonance where
the planets are on horseshoe type orbits with an average eccentricity of about 0.25 to 0.30.
The orbits are aligned and precess slowly around the binary. 

These properties are different from coorbital resonances observed so far. 
In the Saturnian system, Janus and Epimetheus are on nearly circular orbits and in the case of Trojans, very small objects are
in resonance with a much larger planet whose orbit is nearly circular. However, in the latter case the small asteroid can be on an eccentric and inclined
orbit which changes the resonant structure \citep{1999PhRvL..83.2506N}.  
In our case, the resonant planet pair orbits a binary star on quite eccentric orbits.

Having found possible resonant states for only two cases of the Kepler sample, it is presently not known for which orbital parameters of a binary star system
stable coorbital configurations can exits.
Our investigated systems both have very low $e_\mathrm{bin}$, 0.02 and 0.04, which appears to be an important factor for the stability of such systems.
Using a comparable set-up for Kepler-35 ($e_\mathrm{bin}=0.14$), we did not find any coorbital resonant orbits but one of the planets was ejected after an interaction between the two planets. In the case of the highly eccentric binary Kepler-34 with $e_\mathrm{bin}= 0.52$ both planets were ejected after
a close encounter. 

In addition to our equal planet mass studies we ran a model with a different planet-planet mass ratio for the Kepler-47 case.
In this simulation, the mass of the inner planet was about 20\% higher than the outer one with otherwise identical initial conditions.
The system evolved again into a stable coorbital configuration.
However, in this case the planets ended up in a tadpole type orbit around each other with a mean eccentricity of about 0.28.
Further exploration of the parameter space using different binary eccentricities and planet masses will be the topic of future work.

The disc surrounding appears to be essential for the trapping mechanism of the planets.
We also tested the Kepler-47 system with a less viscous disc $\alpha =  10^{-4}$. 
In such a low viscosity disc, the planets are able to open a gap, which leads to joint inward migration where the outer planet pushes the inner planet inward, while 
exciting the eccentricity of the inner planet. Thereby, the inner planet gets ejected as soon as the outer planet migrated closer than two binary separations to the inner planet.
In the case of massive planets, masses $\ge \frac{1}{3}M_\mathrm{Jup}$, both planets reach stable orbits at different semi-major axes. 
Therefore massive planets and inviscid environments may inhibit formation of coorbital resonances. 

Compared to the observations, our planets are located farther away from the stars, which is the consequence of their trapping at the parking position away from the cavity edge. The more massive the planets are, the closer they migrate towards the edge of the cavity.
Also smaller disc viscosities and aspect ratios bring the cavity edge closer to the stars\footnote{These results will be published in a future study.}. However, in such cases, the planets are more vulnerable to the gap-opening which, as explained, can prevent them from getting into 1:1 resonance. Hence, we do not expect the coorbitals form very close to the stars such that the stars' perturbation can affect them. To analyze the general stability of such coorbitals closer to the stars and check if they are similar to single planet orbits, will require extensive future parameter studies for example through n-body simulations.

After having shown that stable coorbital resonances with moderate orbital eccentricities around binary stars are a natural
outcome of a convergent migration process, it will be worthwhile to search for such systems in the Kepler data.
A recent study suggests that the single star TOI-178 is orbited by a coorbital planet pair, and an additional planet \citep{2019A&A...624A..46L}.
\begin{acknowledgements}
We thank Zsolt Sandor for fruitful discussions.
Anna Penzlin was funded by grant KL 650/26-1 of the German Research Foundation
(DFG). 
The authors acknowledge support by the High Performance and
Cloud Computing Group at the Zentrum f\"ur Datenverarbeitung of the University
of T\"ubingen, the state of Baden-W\"urttemberg through bwHPC and the German
Research Foundation (DFG) through grant no INST 37/935-1 FUGG.
All plots in this paper were made with the Python library matplotlib \citep{Hunter:2007}.
\end{acknowledgements}

\bibliography{1to1resonances}
\bibliographystyle{aa}

\end{document}